\documentstyle[12pt,epsfig]{article}
\def\lapproxeq{\lower .7ex\hbox{$\;\stackrel{\textstyle
<}{\sim}\;$}}
\def\gapproxeq{\lower .7ex\hbox{$\;\stackrel{\textstyle
>}{\sim}\;$}}
\begin{document}
\begin{titlepage}
\pagestyle{empty}
\begin{center}
{\large \bf $F_2^{\gamma}$ AT LOW $Q^2$\\
  AND $\sigma_{\gamma \gamma}$ AT HIGH ENERGIES}\footnote{presented by 
  J. Kwieci\'nski at the Cracow Epiphany Conference on 
  electron - positron  colliders,  
  January 1999, Krak\'ow, Poland.}  \\
\vspace{1.1cm}
         {\sc B.~Bade\l{}ek $^a$},  
         {\sc J.~Kwieci\'nski $^b$} and  {\sc A. M.  Sta\'sto $^b$ 
         \footnote{Foundation 
         for Polish Science fellow.}}\\
\vspace{0.3cm}
$^a$ {\it Department of Physics, Uppsala University, P.O.Box 530,
751 21 Uppsala, Sweden} \\
{\it and Institute of Experimental Physics, Warsaw University, Ho\.za 69,
00-681 Warsaw, Poland }\\

$^b$ {\it Department of Theoretical Physics, 
H.~Niewodnicza\'nski Institute of Nuclear Physics, \\
Radzikowskiego 152, 31-342 Cracow, Poland} \\
\end{center}
  \vspace{2cm}
\begin{abstract} 
The parametrisation of the photon structure function in the low $Q^2$ region 
 is formulated.  It includes the VMD contribution and the QCD improved parton 
 model component suitably extrapolated to the low $Q^2$ region.       
The parametrisation describes reasonably well existing experimental data 
on $\sigma_{\gamma \gamma}$ for real photons and the low $Q^2$ data on 
$\sigma_{\gamma^* \gamma}$.    Predictions for 
$\sigma_{\gamma \gamma}$  and for $\sigma_{\gamma^* \gamma}$  
for energies which may be accesssible in future linear colliders 
are also given.
\end{abstract}
\end{titlepage}
The structure function of the photon is described at large scales $Q^2$ by the 
QCD improved parton model \cite{LAC,YELLOW,KRAWZ}.   It is expected however 
that  in the low  $Q^2$ 
region the Vector Meson Dominance (VMD) contribution \cite{VMD} may also become 
 important. 
 Here, as usual, 
$Q^2=-q^2$ where $q$ denotes the four momentum of the virtual photon probing 
the 
real photon with four momentum $p$.  The CM energy squared $W^2$ 
of the $\gamma^* \gamma $ system is  $W^2=(q+p)^2$. \\
In this talk we wish to present the representation of the photon structure 
function which includes both the VMD contribution together with the QCD improved 
parton model term suitably extrapolated to the low $Q^2$ region. This representation 
of the photon structure function is based on the extension of similar representation 
of the nucleon structure function to the case of the photon "target" 
\cite{BBJK1,BBJK2}.  Possible parametrization of the photon structure function 
which extends to the low $Q^2$ region has also been discussed in ref. \cite{GLM}. 
There do also exist several microscopic models describing the energy dependence 
of the total $\gamma \gamma$ cross-section \cite{CGP}. \\
Our representation of the  structure function $F_2(W^2,Q^2)$ is based 
on the following decomposition:  
\begin{equation}  
  F_2(W^2,Q^2) = F_2^{VMD}(W^2,Q^2) + F_2^{partons}(W^2,Q^2)
\label{dec}
\end{equation}
where in what follows we shall consider the structure function of the photon, 
i.e. 
$F_2 \equiv F_2^{\gamma}$.    
The terms $F_2^{VMD}(W^2,Q^2)$ and $F_2^{partons}(W^2,Q^2)$ denote the VMD and 
partonic contributions respectively. The VMD part is given by the following 
formula: 
\begin{equation}   
  F_2^{VMD}(W^2,Q^2)={Q^2\over 4 \pi} \sum_{v=\rho,\omega,\phi} 
   {M_v^4\sigma_{\gamma v}(W^2)\over \gamma_v^2 (Q^2 + M_v^2)^2}
\label{vmd}
\end{equation}
where $M_v$ is the mass of the vector meson $v$ and $\sigma_{\gamma v}(W^2)$  
denotes the $\gamma v$ total cross-section. The parameters $\gamma_v$ can 
be determined from the leptonic widths $\Gamma^v_{e^+e^-}$ \cite{VMD,BBJK1}: 
\begin{equation}
 {\gamma_v^2 \over \pi}= {\alpha^2 M_v\over 3 \Gamma^v_{e^+e^-}}
\label{gammav}
\end{equation}
The partonic contribution 
is expressed  in terms of the structure function 
$F_2^{QCD}$   
obtained from the QCD improved parton model analysis of the photon structure function 
in the large $Q^2$ region 
\cite{BBJK2}:   
\begin{equation}  
  F_2^{partons}(W^2,Q^2) = {Q^2\over Q^2 + Q_0^2} F_2^{QCD}(\bar x,Q^2+Q_0^2) 
\label{partons}
\end{equation} 
where   
\begin{equation}
  \bar x =x \left(1 + {Q_0^2\over Q^2}\right )={Q^2+Q_0^2\over W^2 + Q^2}
\label{xbar}
\end{equation}
with $x$ denoting the Bjorken variable, i.e. $x=Q^2/(2pq)$. The parameter 
$Q_0^2$ 
should have its magnitude greater than the mass squared of the heaviest vector 
meson 
included in the VMD part and its value will be taken to be the same as in 
ref. \cite{BBJK2}, i.e. $Q_0^2=1.2$  GeV$^2$.\\
The $\gamma^* \gamma$ total cross-section $\sigma_{\gamma^* \gamma}$ 
is related in the following way to the photon structure function:   
\begin{equation}
\sigma_{\gamma^* \gamma}(W^2,Q^2)=  {4 \pi^2 \alpha \over Q^2} F_2(W^2,Q^2)
\label{sigma}
\end{equation}
After taking in equation (\ref{sigma}) the limit $Q^2=0$ (for fixed $W$)  we obtain the total  
cross-section $\sigma_{\gamma \gamma}(W^2)$ corresponding to the interaction 
of two real photons.  The representation (\ref{dec}) and equations 
(\ref{vmd}) and (\ref{partons})  give the following 
expression for this cross-section:     

\begin{equation}
  \sigma_{\gamma \gamma}(W^2) = \alpha \pi \sum_{v=\rho,\omega,\phi} 
  {\sigma_{\gamma v}(W^2)\over \gamma_v^2} + {4 \pi^2 \alpha \over Q_0^2} 
  F_2^{QCD}(Q_0^2/W^2,Q_0^2)
\label{sgamgam}
\end{equation}

In the large $Q^2$ region the structure function given by eq. (\ref{dec}) 
becomes equal 
to the QCD improved parton model contribution $F_2^{QCD}(x,Q^2)$.  
The VMD component gives the power correction term which vanishes as ($1/Q^2$) 
for large $Q^2$.  The modifications of the QCD parton model contribution 
(i.e. replacement of the parameter $x$ by $\bar x$ defined by equation 
(\ref{xbar}), the shift of the scale $Q^2 \rightarrow Q^2 + Q^2_0$ and the 
factor $Q^2/(Q^2+Q_0^2)$ instead of 1) are also negligible at large $Q^2$ and introduce 
the power corrections which vanish as $1/Q^2$. \\

In the quantitive analysis of $\sigma_{\gamma* \gamma}$ and of 
$\sigma_{\gamma \gamma}(W^2)$ we have taken the   
  $F_2^{QCD}$ from the LO analysis presented in ref. \cite{GRV}.\\

The VMD part was estimated using the following assumptions: 
\begin{enumerate}
\item The numerical values of the couplings $\gamma_v^2$ are 
the same as those used in ref. \cite{BBJK1}. They 
were estimated from relation (\ref{gammav}) 
which gives the following values : 
\begin{equation}
{\gamma_{\rho}^2\over \pi}=1.98 ~~~~~~~ {\gamma_{\omega}^2\over \pi} = 21.07 
 ~~~~~~~ {\gamma_{\phi}^2\over \pi} = 13.83
 \label{numbers}
 \end{equation}
 
\item The cross-sections   
  $\sigma_{\gamma v}$ are represented as the sum of  the Reggeon and Pomeron 
  contributions: 
\begin{equation}
 \sigma_{\gamma v}(W^2) = R_{\gamma v}(W^2)+P_{\gamma v}(W^2)  
\label{rpgv}
\end{equation} 
where
\begin{equation}
 R_{\gamma v}(W^2) = a^R_{\gamma v}\left({W^2\over W_0^2} \right)^{\lambda_R} 
\label{rgv}
\end{equation}
\begin{equation}
 P_{\gamma v}(W^2) = a^P_{\gamma v}\left({W^2\over W_0^2} \right)^{\lambda_P}
\label{pgv}
\end{equation}
with
\begin{equation} 
\lambda_R=-0.4525, ~~~~~~~~~~~ \lambda_P= 0.0808
\label{lambda}
\end{equation} 
and $W_0^2 = 1 GeV^2$ \cite{DL}. \\

\item The pomeron couplings $a^P_{\gamma v}$ are related to the corresponding 
couplings $a^P_{\gamma p}$ controlling the pomeron contributions to the total $\gamma p$ 
cross-sections assuming the additive quark model and reducing the total 
cross-sections for the interaction of strange quarks by a factor equal 2.  
This gives: 
$$ 
 a^P_{\gamma \rho}= a^P_{\gamma \omega}={2\over 3}a^P_{\gamma p}
$$

\begin{equation}
a^P_{\gamma \phi}={1\over 2}a^P_{\gamma \rho}
\label{pvm}
\end{equation}
\item The reggeon couplings $a^R_{\gamma v}$ are  estimated assuming additive quark
 model and duality (i.e. dominance of planar quark diagrams).  We also assume 
 that the quark couplings to a photon are 
 proportional to the quark charge with the flavour independent proportionality 
 factor.  This gives:  
 $$
 a^R_{\gamma \rho}=a^R_{\gamma \omega}={5\over 9}a^R_{\gamma p}
 $$
 
 \begin{equation}
  a^R_{\gamma \phi}=0
  \label{rvm}
  \end{equation}
\item The couplings $a^P_{\gamma p}$ and  $a^R_{\gamma p}$ are taken from 
the fit discussed in ref. \cite{DL} which gave: 
\begin{equation}   
a^R_{\gamma p} = 0.129 mb,~~~~~~~~~ a^P_{\gamma p} = 0.0677mb 
\label{apr}
\end{equation}

\end{enumerate}

In Fig.1 we compare our predictions  with the data on 
$\sigma_{\gamma \gamma}
(W^2)$ \cite{PLUTO,TPC,MD1,LEP}.  We show experimental points corresponding to 
the "low" energy region ($W<$ 10 GeV) \cite{PLUTO,TPC,MD1}  and  
the recent 
preliminary high energy data 
obtained by the L3 and OPAL collaborations at LEP \cite{LEP}.   
We can see that the representation (\ref{sgamgam}) for 
the total $\gamma \gamma$ cross-section describes the data reasonably well.  
It should 
be stressed that our prediction is essentially parameter free. 
The magnitude of the cross-section is  dominated by the 
VMD component yet the partonic part is also non-negligible.  The latter 
term is in particular responsible for generating steeper increase of the total 
cross-section with increasing $W$ than that embodied in the VMD part which 
is described by the soft pomeron contribution.  The decrease of the 
total cross-section  with increasing energy 
in the low $W$ region is controlled by the Reggeon component of the 
VMD part (see eqs. (\ref{rpgv}), (\ref{rgv}) and (\ref{lambda}))
 and by the valence part of the partonic contribution. 
\begin{figure}[htb]
\centerline{\epsfig{file=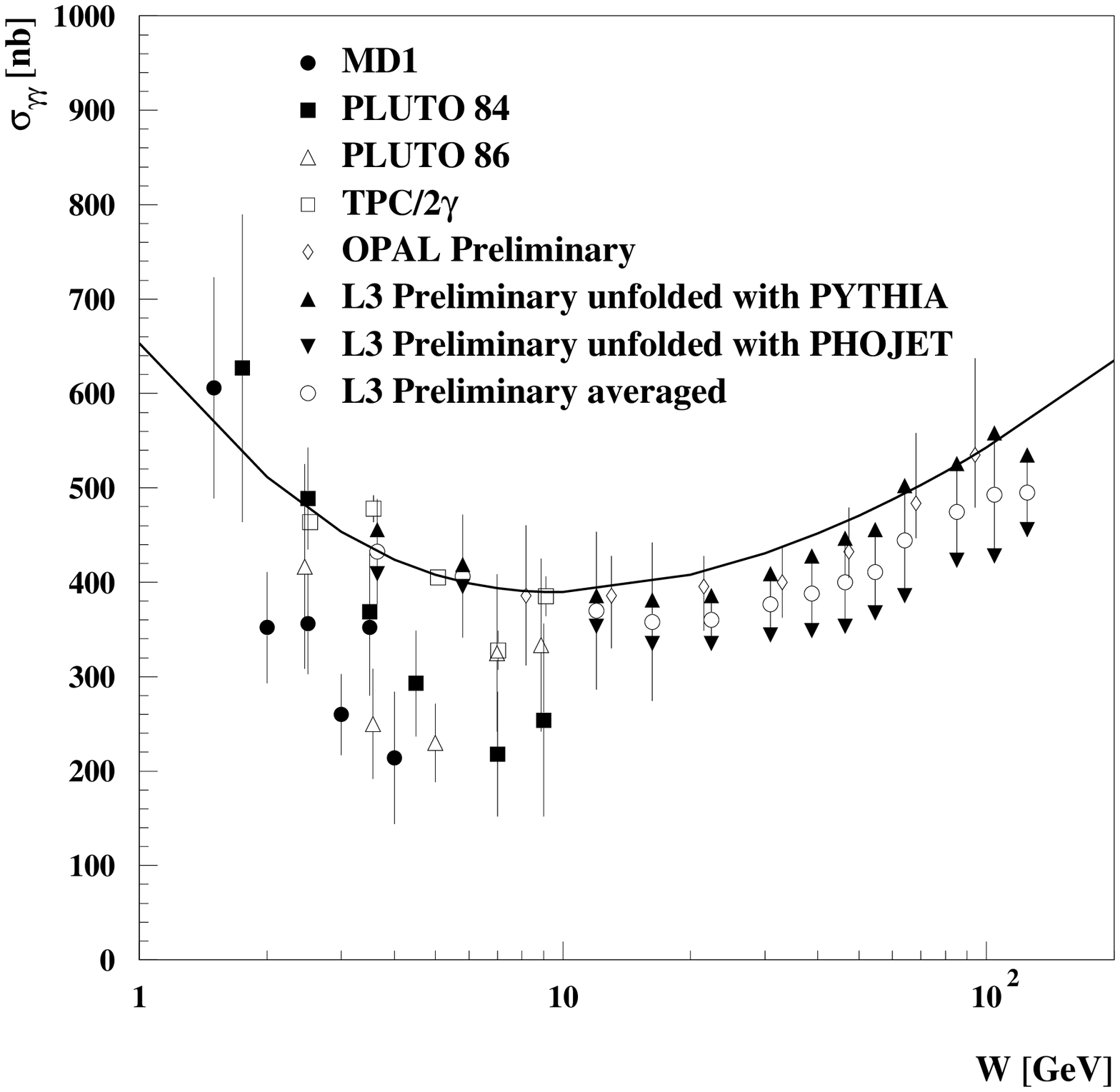,height=12cm,width=12cm}}
\small{Figure 1: Comparison of our predictions for $\sigma_{\gamma \gamma}(W^2)$
based on equation (\ref{sgamgam}) with experimental results
 \cite{PLUTO,TPC,MD1,LEP}}
\label{fig:fig1}
\end{figure} 
 
 In Fig. 2 we show predictions for the total $\gamma \gamma$ cross-section 
 as the function of the total CM energy $W$ in the wide energy range 
 which includes the energies that might be accessible in future 
 linear colliders.  
We also show in this Figure the  decomposition of $\sigma_{\gamma \gamma}
(W^2)$  into its VMD and partonic components.   We see that at very 
high energies these two terms 
exhibit different energy dependence.  The VMD part is described 
by the soft pomeron contribution which gives  the $W^{2 \lambda}$ behaviour 
with $\lambda$ = 0.0808 (\ref{lambda}).  The partonic component increases 
faster with energy since  its energy 
dependence reflects increase of $F_2^{QCD}(\bar x, Q_0^2)$ with decreasing 
$\bar x$ generated by the QCD evolution \cite{GRV}.
\begin{figure}[htb]
\centerline{\epsfig{file=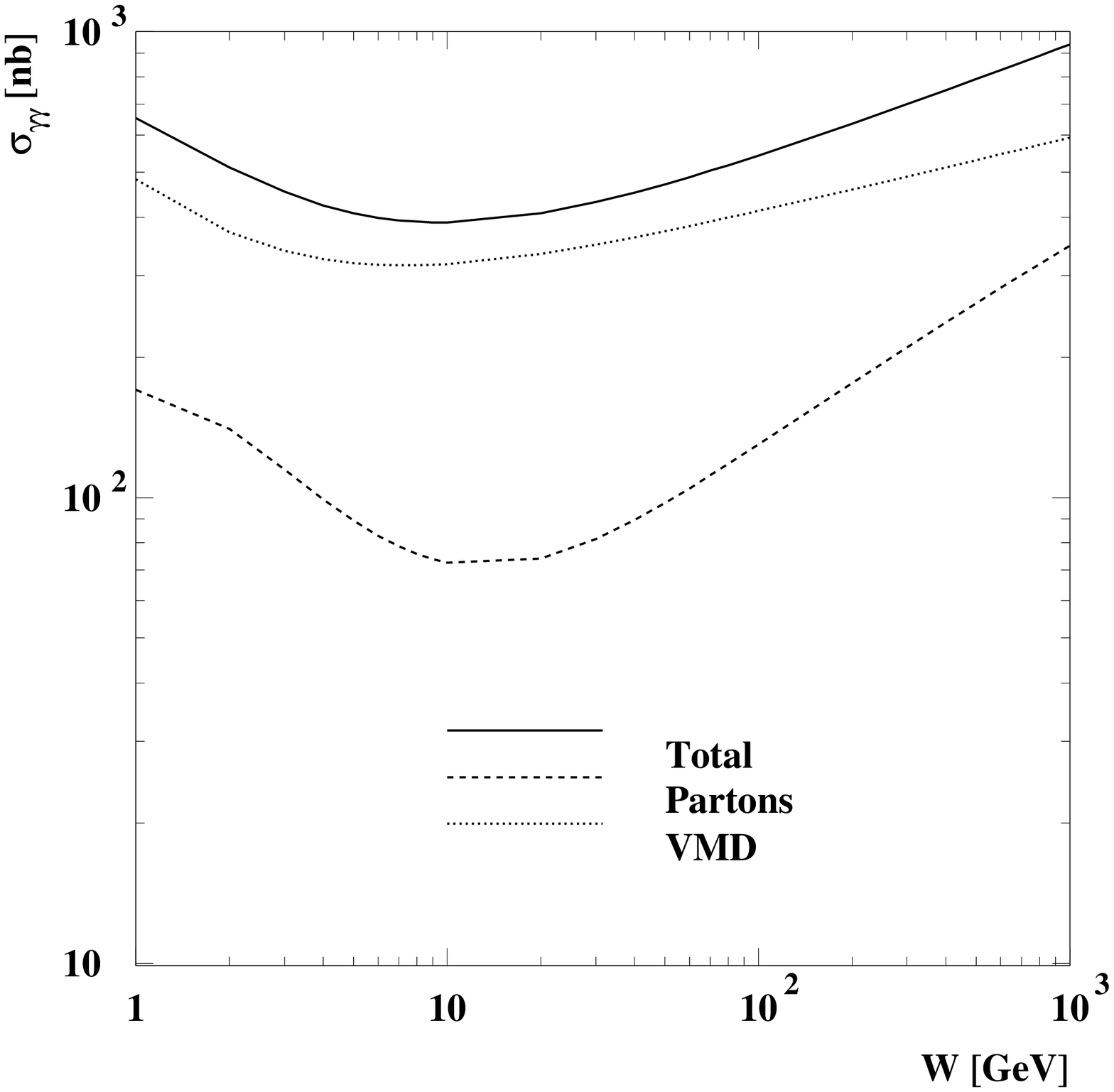,height=10cm,width=10cm}}
\small{Figure 2: The total $\gamma \gamma$ cross-sections $\sigma_{\gamma \gamma}(W^2)$ 
(continuous line)  
calculated from equation (\ref{sgamgam}) and plotted as the function of 
the CM energy $W$  in the wide  energy 
range which includes the region that will be accessible in future linear 
colliders.  We also show separately the VMD (dotted line) and partonic 
(dashed line) components of $\sigma_{\gamma \gamma}(W^2)$.  They correspond  
to the first and the second term on the r.h.s. of equation (\ref{sgamgam}) 
respectively.}
\label{fig:fig2}
\end{figure}  

  This increase is 
stronger than that implied by the soft pomeron exchange.  As the result 
the total $\gamma \gamma$ cross-section, which is the sum of the 
VMD and partonic components does also exhibit stronger increase with 
the increasing energy 
than that of the VMD component.  It is however milder than the 
increase generated by the partonic component alone,  
 at least for $W < 10^3$ GeV. This follows from the fact that  
 in this energy 
 range the magnitude of the cross-section is still dominated by its VMD 
 component.   
We found that for sufficiently high energies $W$ the total $ \gamma 
\gamma$ cross-section 
$\sigma_{\gamma \gamma}
(W^2)$ described by eq. (\ref{sgamgam}) can be parametrized by  the 
effective  power 
law dependence   $\sigma_{\gamma \gamma}
(W^2) \sim (W^2)^{\lambda_{eff}}$ with $\lambda_{eff}$ slowly increasing with 
energy within the range $\lambda_{eff} \sim 0.1 - 0.12$ for 30 GeV $< W <$ 10$^3$ 
GeV.\\
\begin{figure}[htb]
\centerline{\epsfig{file=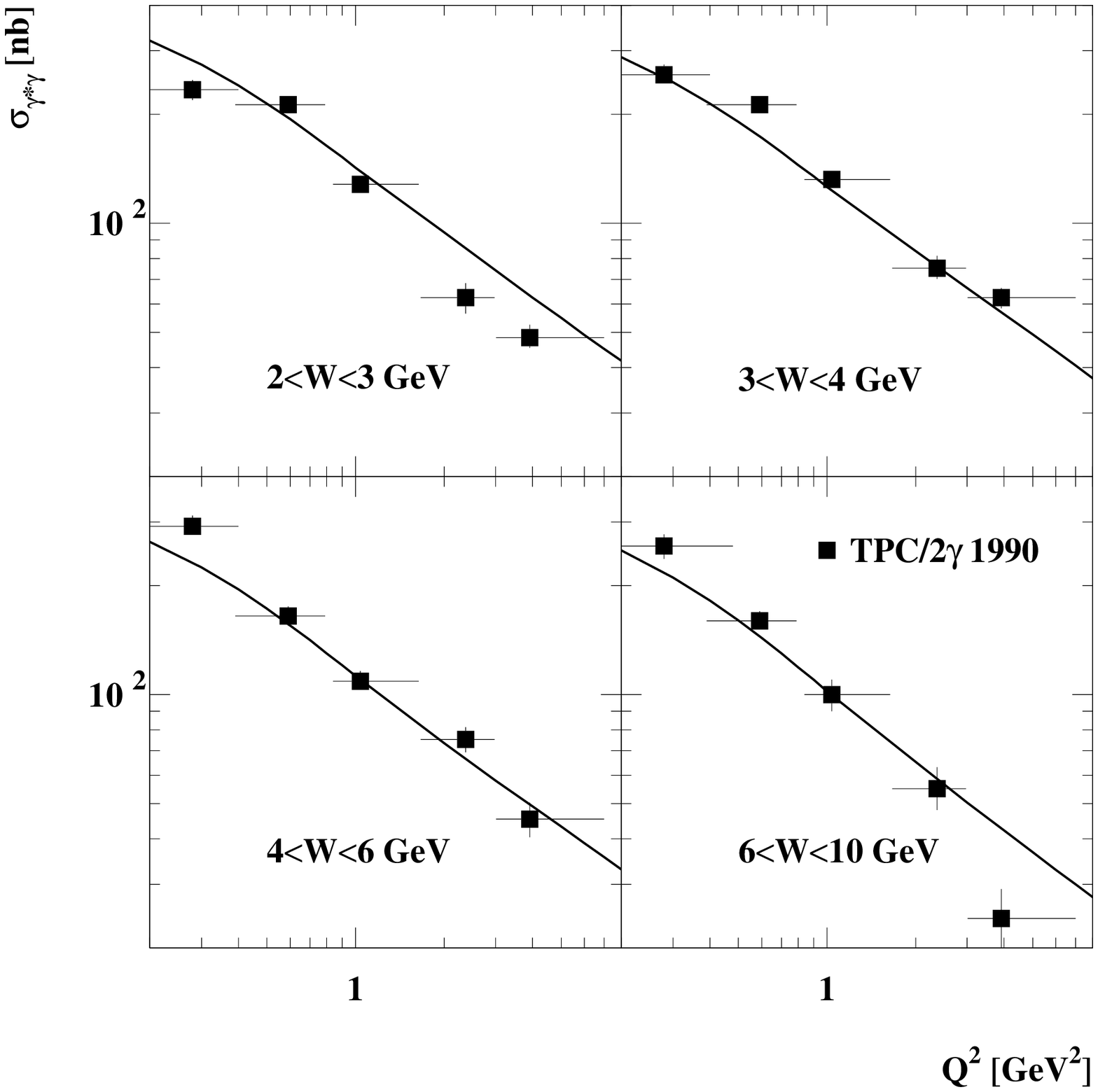,height=12cm,width=12cm}}
\small{Figure 3: Comparison of  predictions for $\sigma_{\gamma* \gamma}(W^2,Q^2)$ 
in the low $Q^2$ region  
based on equations (\ref{dec}, \ref{vmd}, \ref{partons}) and (\ref{sigma})  
with  experimental results \cite{TPC}. }  
\label{fig:fig3}
\end{figure}
In Fig. 3 we compare our predictions for  $\sigma_{\gamma* \gamma}(W^2,Q^2)$ 
based on 
equations(\ref{dec}, \ref{vmd}, \ref{partons}) and (\ref{sigma}) 
with the experimental data in the low $Q^2$ region \cite{TPC}.  We can 
see that in this case the  model is also able to give a good description 
of the data.  \\
Finally in Fig. 4 we show our results for  $\sigma_{\gamma* \gamma}(W^2,Q^2)$ 
plotted as the function of $Q^2$ for different values of the total CM 
energy $W$. 
We notice that for low values $Q^2$ the cross-section does only weakly 
depend 
upon $Q^2$.  In the large $Q^2$ region it follows the 
 $1/Q^2$ scaling behaviour 
modulated by the logarithmic scaling violations implied by perturbative QCD.  \\
\begin{figure}[htb]
\centerline{\epsfig{file=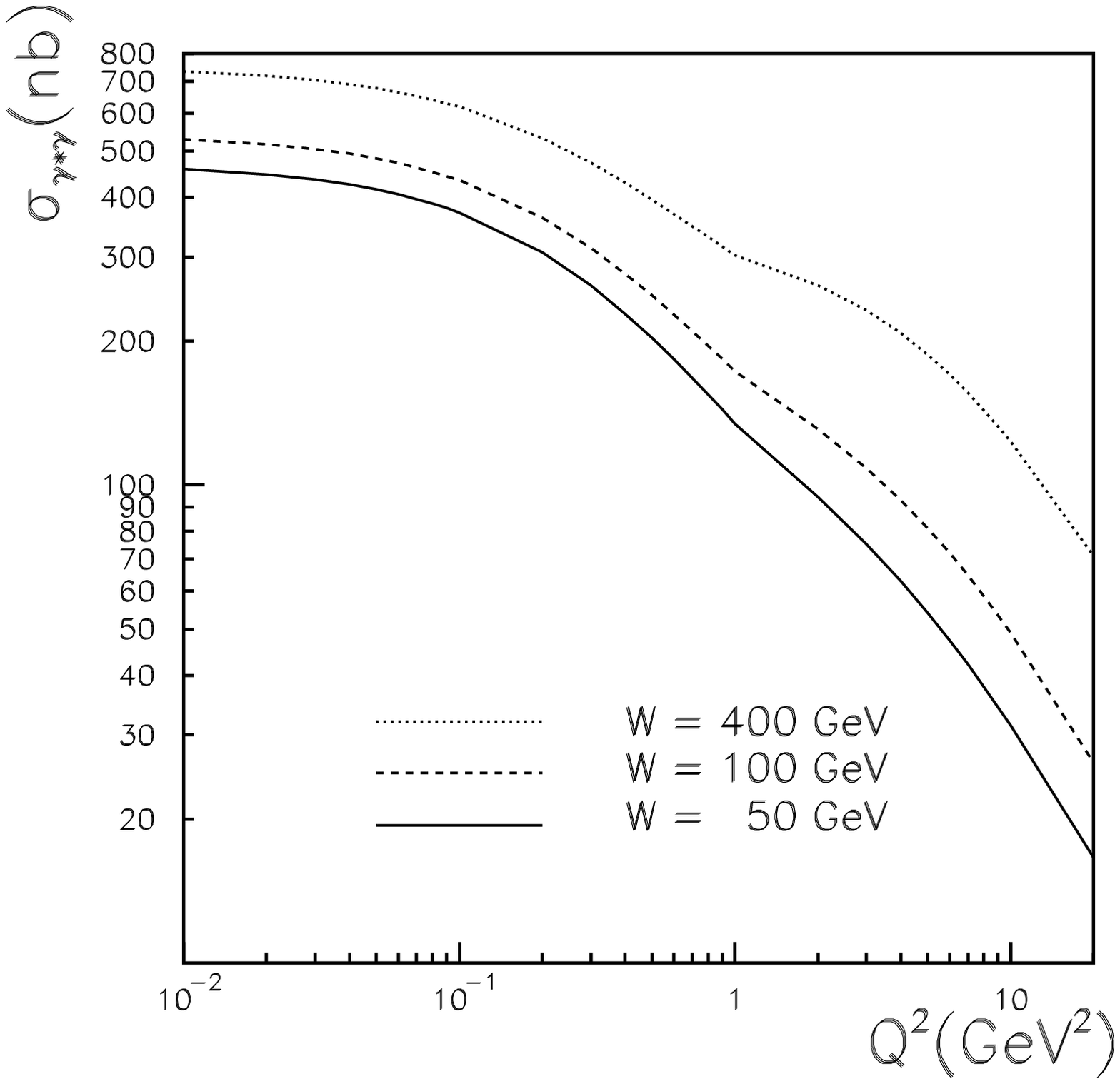,height=12cm,width=12cm}}
\small{Figure 4: Plot of $\sigma_{\gamma* \gamma}(W^2,Q^2)$ 
as the function of $Q^2$ for different values of the CM energy $W$.}  
\label{fig:fig4}
\end{figure}
%
To sum up we have presented  an extension of the representation 
developed  in refs. \cite{BBJK1,BBJK2}    
for the nucleon structure function $F_2$ for arbitrary values of $Q^2$, onto the structure function of the 
real photon .  This representation includes 
both the VMD contribution and the QCD improved parton model component 
suitably extrapolated to the region of low $Q^2$.  We showed that it 
is fairly succesful in describing the experimental data 
on $\sigma_{\gamma \gamma}(W^2)$ and on $\sigma_{\gamma* \gamma}(W^2,Q^2)$ 
at low $Q^2$.  
We also showed that one can naturally explain the fact that the 
increase of the total $\gamma \gamma$ cross-section with increasing CM energy $W$  is stronger than that 
implied by soft pomeron exchange.  The calculated total $\gamma \gamma$ 
cross-section was 
found to exhibit approximate power-law increase with increasing energy $W$, 
i.e. $\sigma_{\gamma \gamma}(W^2) \sim (W^2)^{\lambda_{eff}}$ 
with $\lambda_{eff}$ slowly increasing with 
energy within the range $\lambda_{eff} \sim 0.1 - 0.12$ for 
30 GeV $< W <$ 10$^3$ 
GeV.\\

\section*{Acknowledgments} 
 We congratulate Marek Je\.zabek for organizing an excellent conference.
We thank Maria Krawczyk for several useful discussions.   
This research was partially supported by the Polish State Committee for 
Scientific Research grants 2 P03B 184 10, 
2 P03B 89 13 and by the EU Fourth Framework Programme 
"Training and Mobility of Researchers", 
Network 'Quantum Chromodynamics and the Deep Structure of Elementary 
Particles', contract FMRX - CT98 - 0194.

\end{document}